\documentclass[10pt]{article}

\usepackage[T1]{fontenc}
\usepackage[utf8]{inputenc}
\usepackage{lmodern}
\usepackage{amsmath,amssymb,amsfonts,bm}
\usepackage{booktabs}
\usepackage{graphicx}
\usepackage{microtype}
\usepackage[a4paper,top=18mm,bottom=20mm,left=17mm,right=17mm,columnsep=6mm]{geometry}
\usepackage{authblk}
\usepackage{caption}
\usepackage[hidelinks]{hyperref}

\hypersetup{
  pdftitle={Near-Threshold OPE Constraints and High-Energy Diffractive Dynamics in J/psi Photoproduction},
  pdfauthor={Arkadiy I. Syamtomov},
  pdfsubject={Near-threshold OPE constraints, target-mass corrections, and high-energy diffractive dynamics},
  pdfkeywords={J/psi photoproduction, operator product expansion, target-mass corrections, identifiability, gluon distribution, small-x diffraction, gluon GPDs, near-threshold scattering}
}

\newcommand{\GeV}{\mathrm{GeV}}

\newenvironment{paperabstract}
{\begin{center}\begin{minipage}{0.88\textwidth}\small\noindent\textbf{Abstract.}\quad}
{\end{minipage}\end{center}}

\begin{document}

\title{Near-Threshold OPE Constraints and High-Energy Diffractive Dynamics in $J/\psi$ Photoproduction}
\author{Arkadiy I.\ Syamtomov\thanks{arkady.syamtomov@gmail.com}}
\affil{Bogolyubov Institute for Theoretical Physics, National Academy of Sciences of Ukraine, Kyiv, Ukraine}
\date{\small Published in \textit{Nuclear Physics A} \textbf{1075} (2026) 123481}

\makeatletter
\twocolumn[
\begin{@twocolumnfalse}
\maketitle
\vspace{-1.5em}
\begin{paperabstract}
We study the forward $J/\psi N$ amplitude
$\mathcal{M}_{\psi N}(\lambda)$ using a leading-twist Coulombic
operator-product expansion, exact physical-domain target-mass
resummation, and modern gluon PDFs.  Forward-intercept extrapolations
reported by GlueX and CLAS12 constrain the amplitude modulus within a
vector-meson-dominance normalization.  After fitting an overall scale
and a subtraction constant, the exact-TMC and no-TMC descriptions
differ by only $\Delta\chi^2=0.021$, while a two-parameter empirical
exponential gives an equally good fit.  The point $C_{\rm sub}=0$
changes the profiled minimum by only $0.033$.  Current near-threshold
data therefore determine neither the target-mass correction nor the
subtraction constant separately.  An independent correlated fit to
HERA data gives $\delta=0.695\pm0.023$.  The absence of direct-proton
measurements between $W\simeq4.73$ and $20~\mathrm{GeV}$ prevents a
data-driven determination of the crossover to high-energy diffraction.

\vspace{0.55em}\noindent\textit{Keywords:} $J/\psi$ photoproduction; operator product expansion; target-mass corrections; identifiability; gluon distribution; small-$x$ diffraction; gluon GPDs; near-threshold scattering.
\end{paperabstract}
\vspace{1.0em}
\end{@twocolumnfalse}
]
\makeatother

\section{Introduction}

Heavy quarkonium provides a short-distance probe of gluonic dynamics because
its size is small compared with a typical hadronic scale.  In the Coulombic
heavy-quark limit, the forward quarkonium--nucleon amplitude can be organized
in an operator product expansion (OPE) and related to moments of the gluon
distribution
\cite{Peskin:1979va,Bhanot:1979vb,Kharzeev:1994pz,Syamtomov:1996,Syamtomov:1999}.
The resulting sum-rule construction is naturally a near-threshold
short-distance baseline; evaluating a truncated OPE on the physical cut does
not by itself extend its controlled domain to arbitrarily high energy
\cite{Shifman:1978bx}.

At HERA energies, exclusive $J/\psi$ photoproduction instead exhibits a
power-like rise associated with high-energy diffraction at small gluon
momentum fraction \cite{Donnachie:1998gm,Donnachie:2002en}.  Modern
microscopic descriptions include NLO collinear factorization with low-$x$
resummation, $k_T$ factorization with linear NLO BFKL evolution, and nonlinear
BK/dipole evolution
\cite{Flett:2020jpsi,ArroyoGarcia:2019,Hentschinski:2021,HentschinskiRamirez:2026}.
Proton-target $\gamma p$ cross sections extracted from ALICE ultraperipheral
$p$--Pb collisions and LHCb exclusive $pp$ production extend the empirical
context to the TeV region
\cite{ALICE:2019jpsi,ALICE:2023jpsi,LHCb:2018jpsi,LHCb:2025jpsi}, although
those extractions depend on photon fluxes, absorptive survival corrections,
and, for $pp$, separation of the two photon-emitter branches.

Near threshold, physical photoproduction is intrinsically off-forward.  At
$W=M_\psi+m_N$ the allowed interval collapses to the single value
\[
 t_{\min}=t_{\max}=t_{\rm th}
 =-\frac{m_N M_\psi^2}{m_N+M_\psi}\simeq-2.23~\GeV^2,
\]
placing $t=0$ outside the threshold phase space.  The GlueX and CLAS12 quantities
used below are instead forward intercepts obtained by fitting the measured
finite-$t$ dependence and extrapolating that fit to $t=0$.  The physical
reaction also has large skewness; for threshold $J/\psi$ production one has
approximately $\xi_{\rm th}\sim0.6$, while $\xi\to1$ denotes the formal
heavy-quark, large-skewness limit \cite{GuoJiLiu:2021,GuoJiYuan:2024}.  GPD
analyses can then relate leading moments to gluonic energy--momentum-tensor
form factors.  We formulate the analysis in terms of the forward $J/\psi N$
amplitude, using the extrapolated intercepts as its phenomenological
experimental interface.  Physical finite-$t$ production and the associated
gluon-GPD and energy--momentum-tensor observables require a nonforward
description beyond the present forward OPE construction.

The threshold likelihood uses six intercepts reported by GlueX and CLAS12
\cite{GlueX:2023pev,CLAS12:2026jpsi}.  The full GlueX proton
coverage reaches $W\simeq4.73$~GeV, whereas the highest point in the selected
six-point likelihood is at $W\simeq4.60$~GeV.  The recent Hall-C
$J/\psi$-007 measurement provides complementary differential and integrated
information but does not extend the energy coverage beyond GlueX and is not
inserted into the nominal forward-intercept likelihood \cite{Jpsi007:2026}.
Legacy beryllium and deuterium extractions are not treated as equivalent proton
anchor points \cite{Gittelman:1975ix,Camerini:1975cy}.  At high energy, 51
published H1 and ZEUS elastic total-cross-section measurements are analyzed
with correlated systematic uncertainties
\cite{Chekanov:2002xi,Aktas:2005xu,Alexa:2013xxa}.

We investigate the forward amplitude $\mathcal{M}_{\psi N}(\lambda)$
in two experimentally constrained energy domains.  The analysis addresses
three questions: the constraints imposed by the near-threshold forward-intercept
extrapolations on the target-mass-resummed OPE amplitude; the separate
identifiability of the target-mass correction and subtraction constant; and
the information available on its connection to the high-energy diffractive
regime. We also
compare the OPE curve with an equally flexible empirical threshold benchmark.
The threshold and high-energy domains are analyzed separately, without
imposing a transition model between them.

The gluon dependence is evaluated with four exact LHAPDF identifiers:
\begin{center}
\small
\texttt{ABMP16\_5\_nnlo}, \texttt{MSHT20nnlo\_as118},
\texttt{CT18NNLO},\quad
\texttt{NNPDF40\_nnlo\_as\_01180}.
\end{center}
The corresponding references are Refs.~\cite{Alekhin:2017kpj,Bailey:2020ooq,Hou:2019efy,Ball:2021leu,Buckley:2014ana}.
Their complete Hessian or replica ensembles are propagated according to the
native prescription of each set.  Probabilistic PDF-member errors are kept
separate from variations of the gluon-evolution scale, Coulombic binding
scale, and OPE spectral regulator.

\section{Forward $J/\psi N$ amplitude and OPE baseline}
\label{sec:ope}

The primary theoretical quantity is the forward hadronic amplitude
$\mathcal{M}_{\psi N}(\lambda)$.  To compare it with the forward intercepts
quoted by photoproduction experiments, we introduce the explicitly
extrapolated quantity and its ordinary vector-meson-dominance normalization
\cite{Syamtomov:1999,Petrov:2015vmd},
\begin{equation}
\begin{aligned}
 \mathcal{I}^{\rm ext}_{\gamma N}(W)
 &\equiv \left.\frac{d\sigma_{\gamma N\to\psi N}}{dt}
 \right|_{t=0}^{\rm ext} \\
 &=\frac{3\Gamma(\psi\to e^+e^-)}{16\pi\alpha M_\psi}
 \frac{|\mathcal{M}_{\psi N}(\lambda)|^2}{(s-m_N^2)^2}.
\end{aligned}
 \label{eq:vmd}
\end{equation}
The superscript ``ext'' denotes the forward intercept obtained by
extrapolating the measured finite-$t$ dependence.  Equation~\eqref{eq:vmd}
defines the vector-meson-dominance normalization adopted for the
reduced-amplitude analysis.
Here $W\equiv\sqrt{s}$ and
\begin{equation}
 \lambda=\frac{s-M_\psi^2-m_N^2}{2M_\psi}.
 \label{eq:lambda}
\end{equation}
The variable $\lambda$ is the nucleon energy in the $J/\psi$ rest frame and
the crossing/energy variable of the forward $J/\psi N$ amplitude; at the
physical $J/\psi N$ threshold, $\lambda=m_N$.  Equation~\eqref{eq:vmd} is algebraically equivalent to the standard
expanded flux-factor form because
\[
 [s-(m_N+M_\psi)^2][s-(m_N-M_\psi)^2]
 =4M_\psi^2(\lambda^2-m_N^2).
\]
Using this identity cancels the common threshold factors in the
expanded flux representation and yields Eq.~\eqref{eq:vmd}.  The physical
near-threshold domain remains the finite-$t$ region specified above.  Conversely, the model-dependent amplitude
magnitude constrained by an extrapolated intercept is
\begin{equation}
 |\mathcal{M}_{\psi N}(\lambda)|=(s-m_N^2)
 \left[\frac{16\pi\alpha M_\psi}{3\Gamma(\psi\to e^+e^-)}
 \mathcal{I}^{\rm ext}_{\gamma N}(W)\right]^{1/2},
 \label{eq:amplitude_inversion}
\end{equation}
with cross sections understood in natural units.  At high energy, generalized and multichannel vector-dominance contributions
can modify the relation between photoproduction and the hadronic $J/\psi N$
amplitude \cite{Bauer:1978vmd,Hufner:1998vmd}.
Equations~\eqref{eq:vmd} and \eqref{eq:amplitude_inversion} therefore define
a common reduced-amplitude convention whose hadronic interpretation carries
this additional model dependence.  With this normalization,
$\mathcal{M}_{\psi N}$ and the subtraction constant introduced below are
dimensionless.

The real part follows from the once-subtracted dispersion relation used in the
sum-rule formulation \cite{Syamtomov:1996,Syamtomov:1999},
\begin{equation}
\begin{aligned}
 \operatorname{Re}\mathcal{M}_{\psi N}(\lambda)
 &=C_{\rm sub}+\frac{2\lambda^2}{\pi}\,\mathrm{P.V.}\\
 &\quad\times\int_{m_N}^{\infty}\frac{d\lambda'}
 {\lambda'(\lambda'^2-\lambda^2)}\,
 \operatorname{Im}\mathcal{M}_{\psi N}(\lambda').
\end{aligned}
 \label{eq:dispersion}
\end{equation}
The subtraction constant is defined at the subtraction point,
$C_{\rm sub}\equiv\operatorname{Re}\mathcal{M}_{\psi N}(0)$.
The lower limit $\lambda=m_N$, or $W=M_\psi+m_N$, is the
zero-relative-momentum boundary of the incoming on-shell $J/\psi N$ state. It
marks the beginning of the physical $J/\psi N$ scattering region, but not a
common onset for all hadronic inelastic processes. Each allowed final state has
its own threshold, determined by the masses of the produced hadrons.

To construct the forward $J/\psi N$ amplitude, we first specify the contribution
to its imaginary, or absorptive, part associated with quarkonium dissociation.
In the heavy-quark and Coulombic limits, this contribution is evaluated at
leading order in the QCD multipole expansion, where the compact $c\bar c$ state
couples to the gluon field through the chromoelectric dipole operator. The leading-twist baseline constructed from the Bhanot--Peskin partonic
kernel \cite{Bhanot:1979vb,Arleo:2001mp} is written in the two equivalent
forms
\begin{equation}
\begin{aligned}
 \sigma_0(E)
 &=K_\sigma\int_{\epsilon_0/E}^{1}dx\,
 G_N(x,Q)\,z\phi_0(z)\\
 &=K_\sigma\int_{\epsilon_0/E}^{1}\frac{dx}{x}\,
 \mathcal{G}_N(x,Q)\,z\phi_0(z),
 \qquad z=\frac{xE}{\epsilon_0}.
\end{aligned}
\label{eq:sigma0}
\end{equation}
\begin{equation}
\begin{gathered}
 \mathcal{G}_N(x,Q)\equiv xG_N(x,Q),
 \qquad
 K_\sigma=\frac{4}{3\alpha_s m_Q^2},\\[0.35em]
 \phi_0(z)=\frac{2\pi}{3}\left(\frac{32}{3}\right)^2
 \frac{(z-1)^{3/2}}{z^6}\,\Theta(z-1).
\end{gathered}
\label{eq:phi0}
\end{equation}
Here $x\in(0,1)$ is the proton gluon's light-cone momentum fraction, $Q$ is
the PDF factorization/evolution scale, $G_N(x,Q)$ is the ordinary gluon PDF,
and $\mathcal{G}_N=xG_N$ is its momentum density.  The two lines of
Eq.~\eqref{eq:sigma0} are therefore the same convolution written with two
measures: either the ordinary PDF with $dx$, or the momentum density with
$dx/x$.  Native LHAPDF grids and \texttt{xfxQ} return $\mathcal{G}_N$, which is
inserted directly in the second representation, without an additional division
by $x$.  This also fixes the rank-$n$ moment convention used here,
$A_n=\int_0^1dx\,x^{n-1}G_N(x,Q)=\int_0^1dx\,x^{n-2}\mathcal{G}_N(x,Q)$.  
Accordingly, the $dx/x$ representation of
Ref.~\cite{Syamtomov:2026tmc} is evaluated with the momentum density
$\mathcal{G}_N=xG_N$, using the native LHAPDF output directly.  This PDF
measure affects the numerical convolution while leaving the
trace-resummed two-root mapping unchanged.

With this convention fixed, Eq.~\eqref{eq:sigma0} defines the leading-twist
spectral input used for the near-threshold amplitude.  The energy argument is
$E=\lambda$ in the no-TMC baseline and $E=\lambda r_\pm$ in the two shifted
terms below.  The convolution vanishes for $E\leq\epsilon_0$, while
$xE>\epsilon_0$ is the partonic quarkonium-dissociation condition inside the
Bhanot--Peskin kernel.  This condition should not be identified with the
threshold of a particular hadronic final state: $\sigma_0$ supplies the
corresponding discontinuity of the forward amplitude, not a complete hadronic
inelastic cross section.

Equations~\eqref{eq:sigma0}--\eqref{eq:phi0} combine a leading-order,
leading-twist Bhanot--Peskin coefficient with modern NNLO global-fit gluon
distributions.  We use this construction as a phenomenological sensitivity
baseline rather than a fixed-order matched prediction.  The fitted
coefficient $c_{\rm OPE}$ absorbs an overall normalization change but not a
generic energy-dependent NLO contribution; existing NLO calculations of
quarkonium--hadron dissociation find nontrivial threshold effects
\cite{SongLee:2005}.  Higher-twist operators, higher chromoelectric
multipoles, and non-Coulombic finite-size effects may likewise modify both
the normalization and energy dependence.  The variations studied below
quantify the dependence on the PDF scale, binding scale, and spectral
regulator within this baseline.

Target-mass effects are included through the exact physical-domain result
derived in Ref.~\cite{Syamtomov:2026tmc}. Defining the nucleon velocity in the
$J/\psi$ rest frame as
\begin{equation}
 v=\sqrt{1-\frac{m_N^2}{\lambda^2}},\qquad
 r_\pm=\frac{1\pm v}{2},
 \label{eq:velocity_roots}
\end{equation}
the corrected discontinuity is
\begin{equation}
\begin{aligned}
 \sigma_{\rm TMC}(\lambda)
 &=\frac{1}{v}\Bigl[r_+^2\sigma_0(\lambda r_+)\\
 &\qquad+r_-^2\sigma_0(\lambda r_-)\Bigr].
\end{aligned}
 \label{eq:sigmaTMC}
\end{equation}
Equivalently, using the ordinary gluon PDF explicitly,
\begin{align}
 \sigma_{\rm TMC}(\lambda)
 &=\frac{K_\sigma}{v}\int_0^1 dx\,G_N(x,Q)\nonumber\\
 &\quad\times\Bigl[r_+^2 z_+\phi_0(z_+)
 +r_-^2 z_-\phi_0(z_-)\Bigr],\\
 z_\pm&=\frac{x\lambda r_\pm}{\epsilon_0}.
 \label{eq:sigmaTMC_pdf}
\end{align}
where the step functions in $\phi_0$ enforce $z_\pm>1$. This form makes clear
that no extra inverse power of $x$ is introduced by the target-mass mapping.
The exact resummation involves the two kinematic roots $r_\pm$, and both
contributions are required to obtain the physical-domain leading-twist spectral
cross section. Both terms use the same gluon measure and normalization as
Eq.~\eqref{eq:sigma0}. Here ``exact'' refers to the all-orders resummation of
the kinematic target-mass trace series within the leading-twist Coulombic OPE;
genuine higher-twist operators, higher multipoles, and non-Coulombic finite-size
effects are not included \cite{Syamtomov:2026tmc}. The corresponding
absorptive amplitude is
\begin{align}
 \operatorname{Im}\mathcal{M}_{\psi N}^{\rm TMC}(\lambda)
 &=2M_\psi\sqrt{\lambda^2-m_N^2}\,\sigma_{\rm TMC}(\lambda)\\
 &=2M_\psi\lambda\left[
 r_+^2\sigma_0(\lambda r_+)+r_-^2\sigma_0(\lambda r_-)
 \right].
 \label{eq:ImTMC}
\end{align}
Thus the inverse factor $1/v$ in Eq.~\eqref{eq:sigmaTMC} is cancelled by the
incident flux in Eq.~\eqref{eq:ImTMC}. As $v\to0$, both shifted energies
$\lambda r_\pm$ approach $m_N/2$. Since $m_N/2>\epsilon_0$ for the central
parameters, the Coulombic OPE absorptive amplitude approaches a finite limit at
the physical $J/\psi N$ threshold. The cross section $\sigma_{\rm TMC}$ and the
ratio introduced below retain the explicit $1/v$ factor; it is the product of
that ratio with $v$, and hence the absorptive amplitude, that has a finite
strict-threshold limit.

A finite nonzero discontinuity at the lower end of the cut does not imply a
finite pointwise dispersive real part at the same endpoint.  In the present
once-subtracted representation it generates a logarithmic cusp as
$\lambda\to m_N^+$.  The numerical curves therefore describe the resolved
region above threshold; they are not interpreted as a finite prediction at
the exact point $W=M_\psi+m_N$.

To prevent the leading-twist OPE spectral density of
Ref.~\cite{Syamtomov:2026tmc} from being continued without restriction into
the high-energy region, we multiply the absorptive source entering the
dispersion relation by a compact-support window. The lower source boundary
$W_{\rm src,on}$ is the energy up to which the OPE absorptive source is retained
without modification, while the upper source boundary $W_{\rm src,off}$ is the
energy above which that source is set to zero. Between these boundaries it is
switched off smoothly:
\begin{equation}
 \omega_{\rm OPE}(W)=
 \begin{cases}
 1, & W\leq W_{\rm src,on},\\[1mm]
 \begin{aligned}1-6x_s^5+15x_s^4\\[-0.3em]-10x_s^3\end{aligned},
 & \scriptstyle W_{\rm src,on}<W<W_{\rm src,off},\\[1mm]
 0, & W\geq W_{\rm src,off},
 \end{cases}
 \label{eq:ope_source_weight}
\end{equation}
where, inside the switching interval,
\begin{equation}
 x_s=\frac{\ln(W/W_{\rm src,on})}
 {\ln(W_{\rm src,off}/W_{\rm src,on})}.
 \label{eq:source_window_coordinate}
\end{equation}
We choose $W_{\rm src,on}=4.75$~GeV and $W_{\rm src,off}=8.0$~GeV. The lower boundary lies above the published
GlueX proton coverage, so the OPE source is unchanged throughout the modern
direct-proton threshold region.  The upper boundary limits its contribution
to the dispersive integral.  The compact window regulates the continuation
of the truncated OPE spectral representation; its boundaries are varied in
Sec.~\ref{sec:uncertainties} and are not identified with an observable
crossover scale. With
$W(\lambda')=\sqrt{M_\psi^2+m_N^2+2M_\psi\lambda'}$, the dispersive
contribution is evaluated as
\[
\begin{aligned}
 D[\lambda(W)]
 &=\frac{2\lambda(W)^2}{\pi}\,\mathrm{P.V.}
 \int_{m_N}^{\infty}\frac{d\lambda'}
 {\lambda'(\lambda'^2-\lambda(W)^2)}\\
 &\quad\times\omega_{\rm OPE}[W(\lambda')]
 \operatorname{Im}\mathcal{M}_{\psi N}^{\rm TMC}(\lambda').
\end{aligned}
\]
The source window is evaluated at the integration energy
$W(\lambda')$. The choice of this spectral window and its
sensitivity is discussed in Sec.~\ref{sec:uncertainties}.

For the numerical evaluation, the masses, width, and electromagnetic coupling
in Table~\ref{tab:inputs} follow the Particle Data Group values
\cite{PDG:2024}. The remaining entries list the binding scale, heavy-quark
mass, coupling, and gluon-evolution scale used for the central OPE prediction.
The evolved gluon distributions are taken from the four NNLO sets, while
$\alpha_s=0.3$ is kept fixed in the Coulombic coefficient.

\begin{figure}[t]
 \centering
 \includegraphics[width=0.96\linewidth]{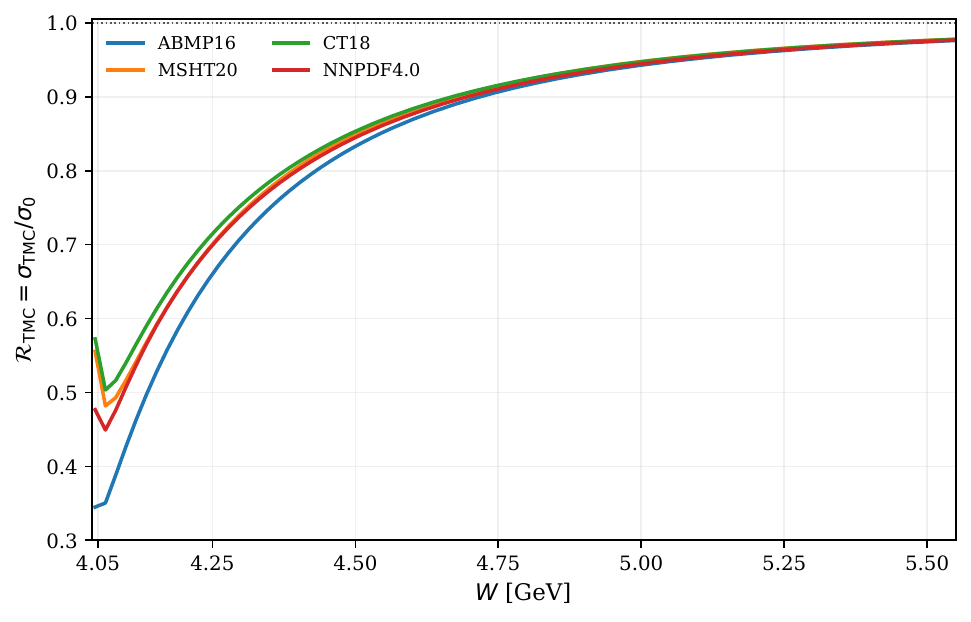}
 \caption{Low-energy view of the exact-TMC to no-TMC ratio for the
forward-amplitude spectral input, for the four central NNLO PDFs.  The ratio characterizes the target-mass modification of the OPE
spectral discontinuity. The displayed range emphasizes the PDF dependence
closest to threshold.}
 \label{fig:rtmc}
\end{figure}

\begin{table}[t]
\centering
\small
\begin{tabular}{lc@{\qquad}lc}
\toprule
quantity & value & quantity & value\\
\midrule
$m_N$ & $0.938272$~GeV & $M_\psi$ & $3.096900$~GeV\\
$\Gamma_{ee}$ & $5.55$~keV & $\alpha$ & $1/137.036$\\
$\epsilon_0$ & $0.16$~GeV & $m_Q$ & $1.5$~GeV\\
$\alpha_s$ & $0.30$ & $Q$ & $10$~GeV\\
\bottomrule
\end{tabular}
\caption{Central numerical inputs.}
\label{tab:inputs}
\end{table}
To illustrate the size and energy dependence of the exact target-mass
correction, Fig.~\ref{fig:rtmc} compares the corrected and uncorrected OPE
predictions through
\begin{equation}
 \mathcal{R}_{\rm TMC}(W)=\frac{\sigma_{\rm TMC}(W)}{\sigma_0(W)}.
 \label{eq:rtmc}
\end{equation}
Here $\sigma_0(W)\equiv\sigma_0[\lambda(W)]$ and
$\sigma_{\rm TMC}(W)\equiv\sigma_{\rm TMC}[\lambda(W)]$. The ratio
$\mathcal{R}_{\rm TMC}$ is therefore used as an above-threshold
diagnostic rather than as a finite observable at the strict threshold. The
first displayed point, $W=4.045$~GeV, lies about $10$~MeV above the physical
$J/\psi N$ threshold. At this energy the partonic dissociation
condition restricts the convolution to a limited interval of gluon momentum
fractions. Since the integral samples each gluon distribution over only this
narrow $x$ range, the predicted correction differs appreciably among the PDF
sets. The dependence decreases rapidly with energy:
$\mathcal{R}_{\rm TMC}\simeq0.945$--$0.950$ near $5$~GeV,
$0.9868$--$0.9876$ near $6$~GeV, and approximately $0.9994$ near $10$~GeV.
The correction falls below one percent for all four central sets above about
$6.4$~GeV.

Representative numerical values of $\mathcal{R}_{\rm TMC}$ and of the finite
flux-weighted product $v\mathcal{R}_{\rm TMC}$ are collected in
Table~\ref{tab:tmc}.

\begin{table*}[t]
\centering
\small
\begin{tabular}{ccc}
\toprule
$W$ [GeV] & four-PDF range of $\mathcal{R}_{\rm TMC}$ & $v\mathcal{R}_{\rm TMC}$\\
\midrule
4.0450  & 0.345--0.572 & 0.056--0.094\\
4.1025  & 0.434--0.545 & 0.176--0.221\\
4.5050  & 0.835--0.855 & 0.673--0.689\\
5.0225  & 0.9452--0.9498 & 0.8688--0.8730\\
6.0000  & 0.98675--0.98757 & 0.96085--0.96165\\
10.0750 & 0.999395--0.999417 & 0.997356--0.997378\\
\bottomrule
\end{tabular}
\caption{Representative exact-TMC ratios and finite flux-weighted products.}
\label{tab:tmc}
\end{table*}

\section{Near-threshold amplitude constraints}
\label{sec:threshold_constraints}

The threshold likelihood is constructed from the six forward intercepts
reported by GlueX and CLAS12, obtained by extrapolating their measured
finite-$t$ distributions. Equation~\eqref{eq:vmd} maps each model amplitude to this
extrapolated-intercept representation.  The intercepts constrain the
amplitude modulus within the VMD and finite-$t$ extrapolation assumptions,
while the dispersive OPE construction determines its complex phase.

A correlated systematic uncertainty is represented by a dimensionless
nuisance parameter $\beta_k$.  To allow both absolute and multiplicative
published responses, the profile likelihood is written as
\begin{equation}
 \chi^2(\bm\theta,\bm\beta)=
 \sum_i\frac{\left[d_i-T_i(\bm\theta)-
 \sum_k A_{ik}(\bm\theta)\beta_k\right]^2}{s_i^2}
 +\sum_k\beta_k^2.
 \label{eq:chi2}
\end{equation}
Here $s_i$ contains statistical and genuinely uncorrelated errors, while
$A_{ik}$ is the signed one-standard-deviation response of point $i$ to source
$k$.  For a relative normalization source,
$A_{ik}=r_{ik}T_i(\bm\theta)$; published absolute responses are used directly.
At each $\bm\theta$, the nuisance parameters are profiled by minimizing
Eq.~\eqref{eq:chi2}.

For GlueX and CLAS12, one common scale nuisance is used for each experiment
\cite{GlueX:2023pev,CLAS12:2026jpsi}.  The pointwise CLAS12 uncertainties on
the extrapolated forward intercepts are statistical; a conservative variant additionally includes the
quoted bin-dependent systematic components.  CLAS12 publishes finite photon-energy intervals rather than an effective energy for each intercept.  The
nominal assignment is the arithmetic mean of the point-level mean photon
energies entering the corresponding fit; bin midpoints are retained as a
sensitivity variant.  The adopted numerical inputs are provided in the supplementary tables.

The forward OPE amplitude is parameterized as
\begin{equation}
 \mathcal{M}_{\psi N}^{\rm OPE}(\lambda)=
 \sqrt{N_{\rm OPE}}\,[C_{\rm sub}+D(\lambda)+iI(\lambda)],
 \label{eq:MOPE_physical}
\end{equation}
where $I(\lambda)$ follows from Eq.~\eqref{eq:ImTMC}, and $D(\lambda)$ is the dispersive
contribution generated by the compact OPE source in
Eq.~\eqref{eq:ope_source_weight}.  For numerical minimization we use
\begin{equation}
 c_{\rm OPE}=\sqrt{N_{\rm OPE}},\qquad
 a_{\rm sub}=c_{\rm OPE}C_{\rm sub},
 \label{eq:transform}
\end{equation}
so that
\begin{equation}
 \mathcal{M}_{\psi N}^{\rm OPE}(\lambda)=a_{\rm sub}+c_{\rm OPE}[D(\lambda)+iI(\lambda)].
 \label{eq:MOPE_fit}
\end{equation}
Because both amplitude coordinates are fitted, this likelihood tests the
OPE-generated energy dependence rather than an absolute Coulombic
normalization.

For ABMP16 the exact-TMC fit gives
\begin{equation}
 \chi^2=2.6425,\qquad \mathrm{ndf}=4,\qquad p=0.6193,
 \label{eq:threshold_fit_result}
\end{equation}
with common-scale pulls $\beta_{\rm GlueX}=-0.689$ and
$\beta_{\rm CLAS12}=+0.330$.  The four central PDF results are collected in
Table~\ref{tab:thresholdfit}.

\begin{table*}[t]
\centering
\small
\begin{tabular}{lrrrrr}
\toprule
PDF & $a_{\rm sub}$ & $c_{\rm OPE}$ & $C_{\rm sub}$ & $N_{\rm OPE}$ & $\chi^2$\\
\midrule
ABMP16   & $-0.5642$ & 0.146688 & $-3.85$ & 0.021517 & 2.643\\
MSHT20   & $-1.0121$ & 0.145180 & $-6.97$ & 0.021077 & 2.659\\
CT18     & $-1.0903$ & 0.145825 & $-7.48$ & 0.021265 & 2.650\\
NNPDF4.0 & $-0.8281$ & 0.149499 & $-5.54$ & 0.022350 & 2.634\\
\bottomrule
\end{tabular}
\caption{Exact-TMC fits to the six GlueX and CLAS12 extrapolated forward
intercepts.}
\label{tab:thresholdfit}
\end{table*}

The present sample provides no statistical preference among the tested
threshold shapes.
The independently refitted no-TMC calculation gives $\chi^2=2.6215$, only
$0.0211$ below the exact-TMC result.  As a simple phenomenological shape benchmark,
we also fit
\begin{equation}
 \mathcal{I}_{\gamma p}^{\rm ext,emp}(W)
 =A\exp[b(W-W_\star)],\qquad W_\star=4.40~\GeV .
 \label{eq:empirical_threshold}
\end{equation}
with the same two experiment-scale nuisances.  The fit gives
$A=2.716~\mathrm{nb}~\GeV^{-2}$, $b=3.090~\GeV^{-1}$, and
$\chi^2=2.429$.  All three descriptions contain two unconstrained physical
parameters, so their AIC and BIC differences equal their small
$\chi^2$ differences up to a common constant.  The resulting model comparison is summarized in
Table~\ref{tab:model_compare}.

\begin{table*}[t]
\centering
\small
\begin{tabular}{lccc}
\toprule
Threshold description & $\chi^2$ & $p$ ($\mathrm{ndf}=4$) & $\Delta\mathrm{AIC}$\\
\midrule
Empirical exponential & 2.429 & 0.657 & 0\\
No-TMC OPE refit      & 2.621 & 0.623 & 0.192\\
Exact-TMC OPE refit   & 2.643 & 0.619 & 0.214\\
\bottomrule
\end{tabular}
\caption{Comparison of the equally flexible threshold descriptions.}
\label{tab:model_compare}
\end{table*}

\begin{figure}[t]
 \centering
 \includegraphics[width=0.96\linewidth]{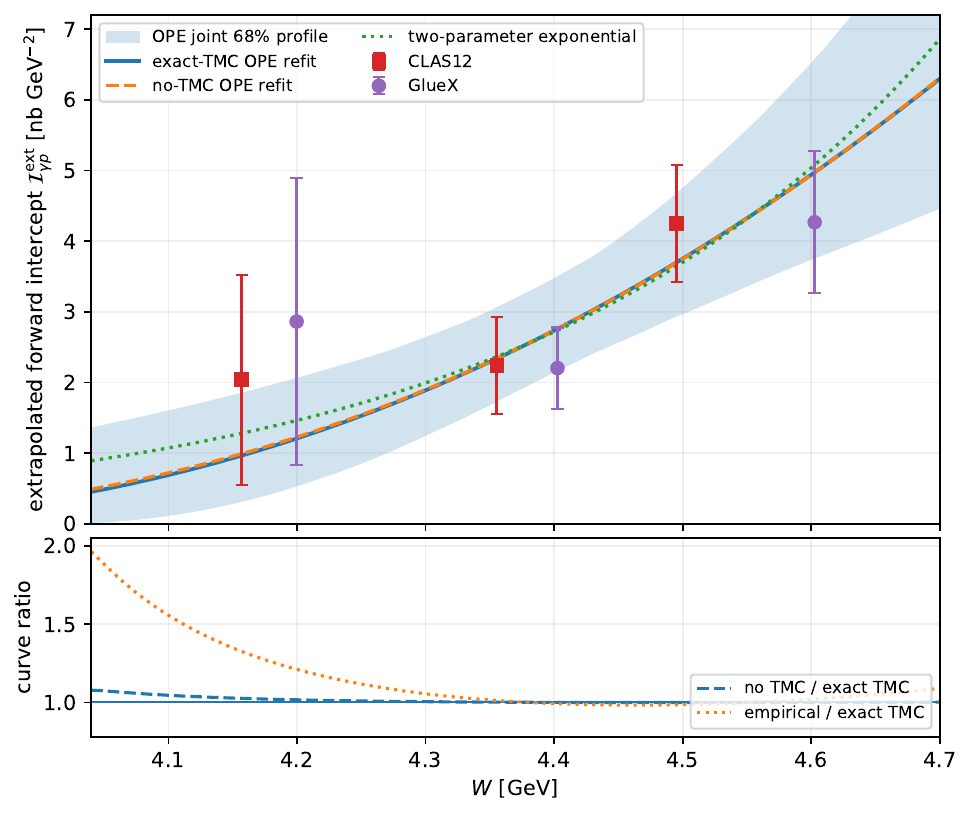}
 \caption{Threshold identifiability in the extrapolated-intercept
representation.  Upper panel: the ABMP16 exact-TMC OPE fit and joint
two-parameter 68\% profile band, compared with the independently refitted
no-TMC OPE curve and a two-parameter empirical exponential.  The GlueX and CLAS12 points are the published forward-intercept
extrapolations obtained from finite-$t$ data. Lower panel: ratios to the
exact-TMC central fit.  The larger separation below the first measurement is
an unconstrained extrapolation.}
 \label{fig:threshold_models}
\end{figure}

Figure~\ref{fig:threshold_models} displays the exact-TMC OPE fit, its joint
profile band, the independently refitted no-TMC curve, and the empirical
benchmark.  Their differences are much smaller than the current experimental
profile uncertainty over the measured interval.  Below the first measured
energy, their separation instead illustrates the unconstrained nature of the
extrapolation.

The subtraction constant is likewise not separately identified.  For ABMP16,
the local Gaussian estimate is $C_{\rm sub}=-3.85\pm19.92$, while the full
one-parameter profile permits approximately $-22<C_{\rm sub}<20$ at 68\% and
$-36<C_{\rm sub}<57$ at 95\%.  The local correlation with $N_{\rm OPE}$ is
$-0.925$, and fixing $C_{\rm sub}=0$ changes the minimum by only
\begin{equation}
 \Delta\chi^2(C_{\rm sub}=0)=0.033.
\end{equation}
The fitted intercept curve therefore constrains the model-dependent
forward-amplitude modulus, whereas $C_{\rm sub}$ and the target-mass
correction remain individually unresolved.

\begin{figure}[t]
 \centering
 \includegraphics[width=0.96\linewidth]{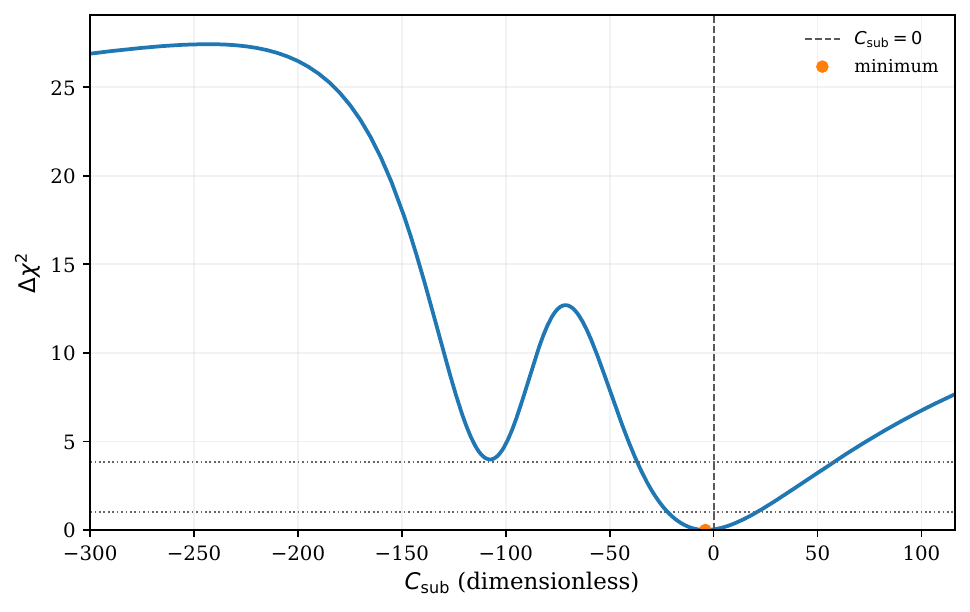}
 \caption{Profile likelihood of the dimensionless ABMP16 subtraction
 constant.  The dotted lines at $\Delta\chi^2=1$ and $3.84$ mark the
 conventional 68\% and 95\% one-parameter levels.}
 \label{fig:csubprofile}
\end{figure}
\section{High-energy benchmark and crossover constraints}
\label{sec:high_energy}

The high-energy likelihood is constructed independently of the threshold fit.
For H1, the published signed decomposition into uncorrelated and 19 correlated
sources is used directly \cite{Alexa:2013xxa}.  ZEUS publishes pointwise
errors and the sizes of leading systematic sources but not a signed bin-by-bin
covariance \cite{Chekanov:2002xi}.  The nominal ZEUS model therefore assigns
one common normalization nuisance to each decay-mode sample and one
energy-dependent tilt calibrated to the published uncertainty on the exponent;
diagonal and maximally correlated alternatives are retained as sensitivity
tests.  Thus the analysis uses the published H1 decomposition and a
reconstructed, explicitly varied ZEUS covariance model.

The 51 HERA elastic total cross sections are fitted with
\begin{equation}
 \sigma_{\rm R}(W)=N_{90}
 \left(\frac{W}{90~\GeV}\right)^\delta,
 \label{eq:hera_power}
\end{equation}
which gives
\begin{equation}
 N_{90}=76.6\pm2.3~\mathrm{nb},\qquad
 \delta=0.695\pm0.023,
 \label{eq:hera_result}
\end{equation}
with correlation $\rho(N_{90},\delta)=-0.072$.  The nominal value is
$\chi^2=26.34$ for 49 degrees of freedom.  The large value $p=0.997$ reflects the conservative point uncertainties
and their strong correlated components, and hence provides limited
discrimination beyond the fitted power-law trend.  Across the principal ZEUS covariance reconstructions,
$\delta$ remains in $0.6911$--$0.7002$.  H1-only and ZEUS-only fits give
$0.690\pm0.030$ and $0.699\pm0.037$, respectively, while removing the H1 2013
high-energy or H1 2006 subset shifts the combined central value by $+0.025$ or
$-0.027$. 

\begin{figure}[t]
 \centering
 \includegraphics[width=0.96\linewidth]{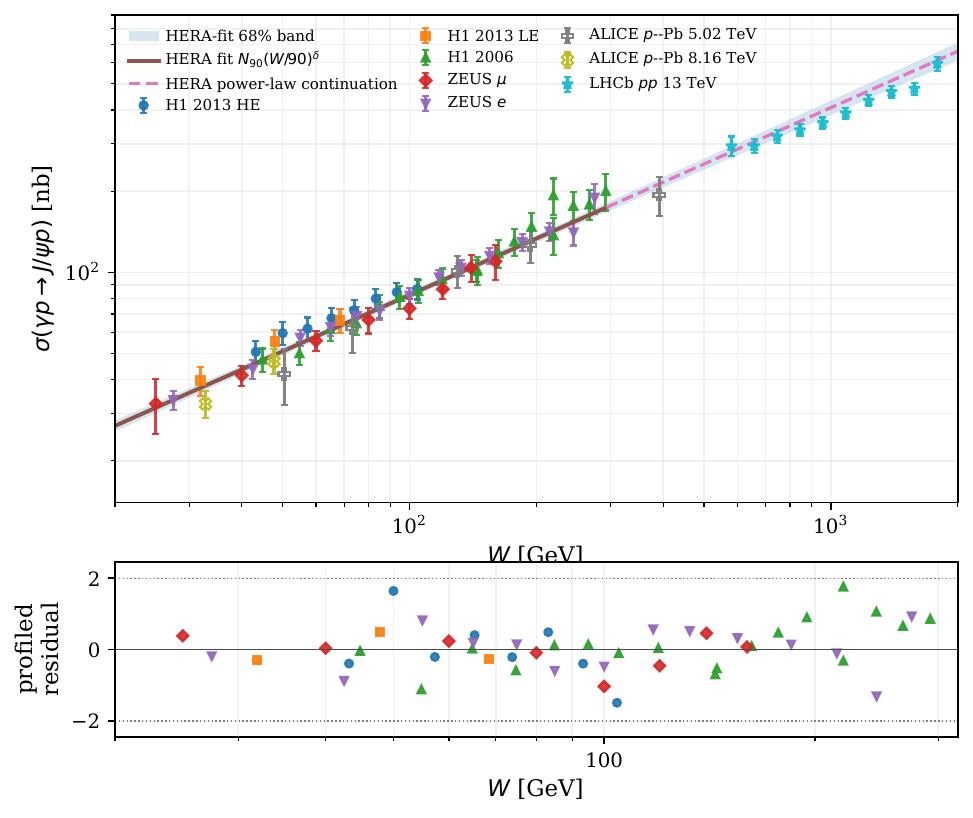}
 \caption{High-energy context.  Upper panel: fit to the H1 and ZEUS elastic
 total cross sections using the published H1 systematic decomposition and the
 nominal reconstructed ZEUS covariance model.  Theory-assisted proton-target
 $\gamma p$ extractions from ALICE and LHCb are shown for comparison but do not
 enter the likelihood.  Lower panel: profiled residuals of the fitted HERA
 points.}
 \label{fig:hera}
\end{figure}
The ALICE and LHCb extractions extend the visual comparison to
$W\simeq1.8$~TeV and are broadly compatible with the HERA power-law
continuation
\cite{ALICE:2019jpsi,ALICE:2023jpsi,LHCb:2025jpsi}.  They are shown as high-energy context rather than included in the
likelihood, because a covariance-consistent combination would require
propagation of photon-flux and survival-factor uncertainties and, for LHCb,
the theory-assisted separation of the two photon-emitter branches.
Agreement with one empirical power does not select among NLO collinear,
linear NLO-BFKL, or nonlinear BK/dipole dynamics.

\begin{figure}[t]
 \centering
 \includegraphics[width=0.96\linewidth]{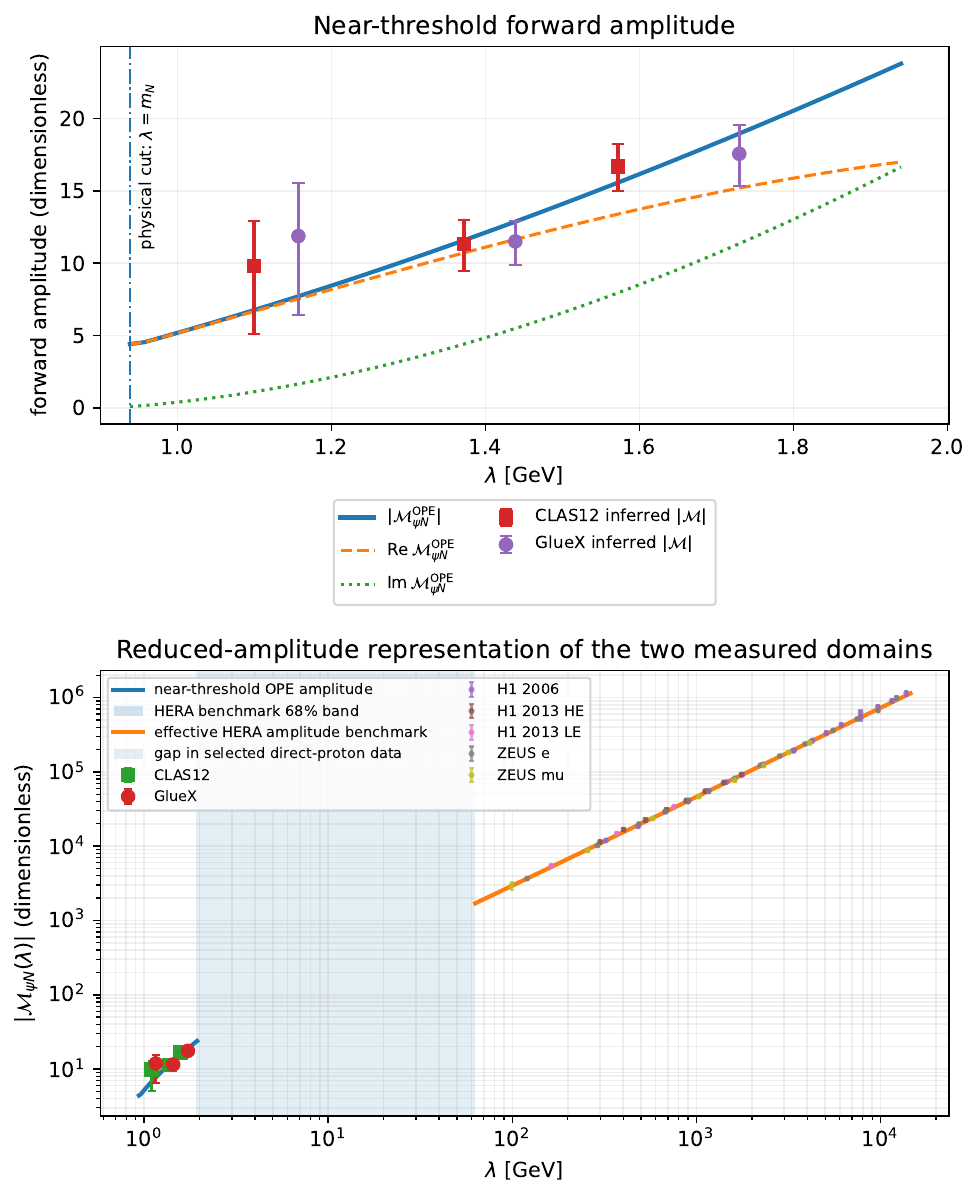}
 \caption{Reduced-amplitude representation of the forward
$J/\psi N$ amplitude.  Upper panel: the real part, imaginary part, and modulus
 of the fitted ABMP16 near-threshold OPE amplitude, with $\lambda=m_N$ marking
 the beginning of the physical $J/\psi N$ cut; the points are modulus values
 inferred from the GlueX and CLAS12 extrapolated intercepts through
 Eq.~\eqref{eq:amplitude_inversion}.  Lower panel: the corresponding modulus in
 the two measured domains.  The HERA benchmark uses the slope conversion of
 Eq.~\eqref{eq:bslope} followed by the same VMD normalization.  No amplitude
 interpolation, transition scale, or phase evolution is inferred in the shaded
 data gap.  Error bars propagate the quoted experimental uncertainties only;
 slope- and VMD-model uncertainties are not included.}
 \label{fig:domains}
\end{figure}

For a common visual normalization only, the HERA total cross section is
converted to a forward intercept using
\begin{equation}
 \begin{aligned}
 b(W)&=b_{90}+4\alpha'\ln\left(\frac{W}{90~\GeV}\right),\\
 b_{90}&=4.5803\pm0.0827~\GeV^{-2},\\
 \alpha'&=0.1236\pm0.0195~\GeV^{-2}.
 \end{aligned}
 \label{eq:bslope}
\end{equation}
Together with Eq.~\eqref{eq:hera_result}, the fitted slope is numerically
compatible with the effective soft-Pomeron/vector-dominance phenomenology
discussed in Ref.~\cite{Petrov:2015vmd}.  The same energy dependence can
also arise in NLO collinear, BFKL, and dipole-based descriptions, so the
effective power and slope do not discriminate among these microscopic
mechanisms.  
For the reduced-amplitude representation in
Fig.~\ref{fig:domains}, the HERA total cross section is first converted to
the effective forward intercept
$\mathcal{I}_{\gamma p}^{\rm ext}=b(W)\sigma_{\rm R}(W)$ and then mapped
through Eq.~\eqref{eq:amplitude_inversion}.  The resulting high-energy curve is an effective reduced-amplitude
benchmark whose interpretation inherits the slope and VMD model
dependence.  The near-threshold
points in the figure are mapped in the same way from the published extrapolated
intercepts.  Photoproduction constrains only the modulus in this construction,
whereas the real and imaginary parts shown in the upper panel follow from the
OPE and dispersion-relation assumptions.

The threshold and HERA likelihoods contain disjoint physical and nuisance
parameters, and no selected measurement lies between them.  Their joint likelihood therefore factorizes;
any smooth path placed in the gap has the same likelihood.  In particular, the
data determine neither a crossover location nor its width or complex phase.
The compact source window in Sec.~\ref{sec:ope} is a regulator of the truncated
OPE dispersive source, not a fitted boundary between separately observable
components.  A field-theoretic matching formula would require a demonstrated
common expansion or overlap term between the local OPE and a small-$x$
amplitude; none is assumed here.
\section{Uncertainties, sensitivity, and numerical checks}
\label{sec:uncertainties}

The threshold likelihood is refitted for every official PDF member.  The scan
contains 255 fits: 30 ABMP16 members, 65 MSHT20 members, 59 CT18 members, and
100 NNPDF replicas plus the separate NNPDF central member.  ABMP16 uses its
symmetric-Hessian prescription; MSHT20 uses 32 paired eigenvectors at 68\%
confidence; the CT18 29-pair 90\% Hessian error is divided by 1.645; and
NNPDF4.0 uses replica statistics.  All member fits converge.

Table~\ref{tab:pdfunc} gives the internal PDF errors of the fitted amplitude
coordinates.  The corresponding uncertainty of the refitted extrapolated intercept at
the six selected energies is below 0.9\% for every set, with a maximum of
0.837\%.  This value quantifies the conditional member-to-member variation after
refitting; the amplitude parameters $a_{\rm sub}$ and $c_{\rm OPE}$ absorb
most PDF-induced changes over the narrow measured interval.  
The quoted PDF-member variation is separate from the perturbative-order
mismatch between the leading-order coefficient and NNLO PDF evolution.

\begin{table*}[t]
\centering
\small
\begin{tabular}{lcccc}
\toprule
PDF & $C_{\rm sub}$ & $\Delta_{\rm PDF}C_{\rm sub}$ & $N_{\rm OPE}$ & $\Delta_{\rm PDF}N_{\rm OPE}$\\
\midrule
ABMP16   & $-3.847$ & 1.390 & 0.021517 & 0.000437\\
MSHT20   & $-6.971$ & 0.613 & 0.021077 & 0.000739\\
CT18     & $-7.477$ & 1.176 & 0.021265 & 0.001080\\
NNPDF4.0 & $-5.539$ & 0.441 & 0.022350 & 0.000427\\
\bottomrule
\end{tabular}
\caption{Internal PDF-member uncertainties after refitting the threshold
amplitude.}
\label{tab:pdfunc}
\end{table*}

\begin{figure}[t]
 \centering
 \includegraphics[width=0.96\linewidth]{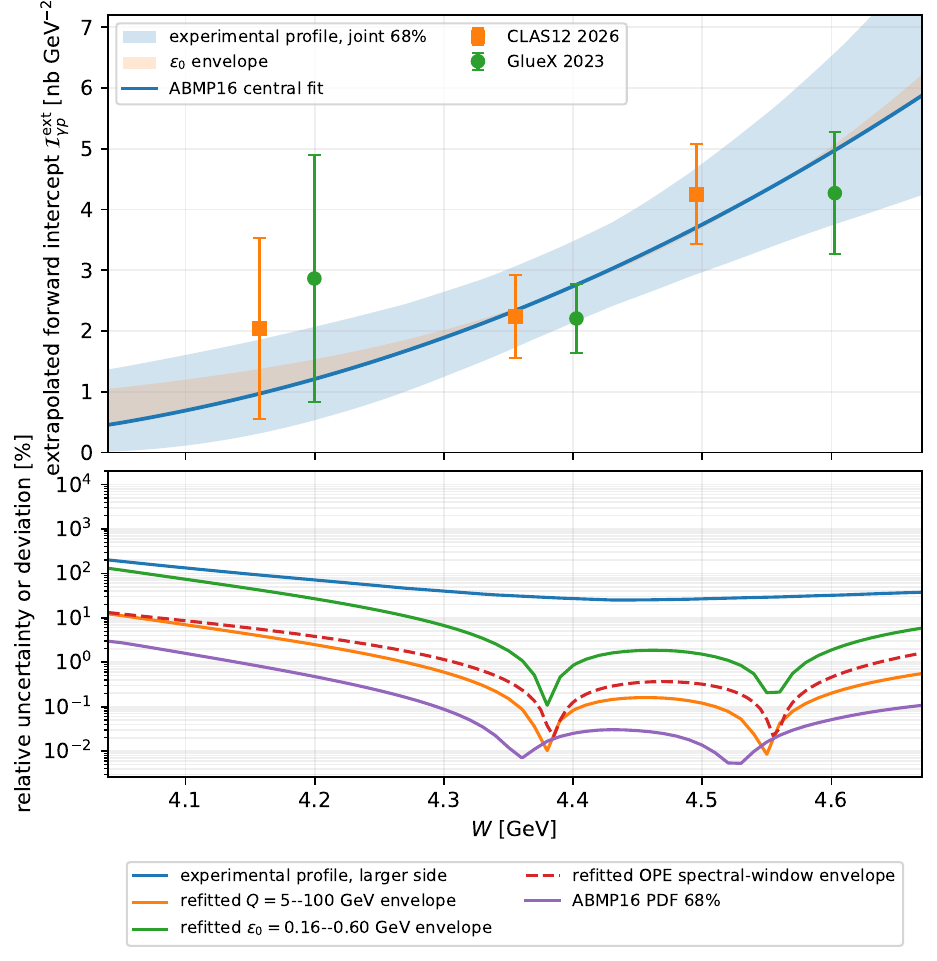}
 \caption{Extrapolated-intercept fit and separately refitted
experimental-profile, PDF-member, PDF-scale, Coulombic binding-scale, and OPE
source-regulator variations.  The GlueX and CLAS12 points are extrapolated
forward intercepts.  The discrete model envelopes are reported separately from the
probabilistic experimental and PDF uncertainties.}
 \label{fig:uncertainties}
\end{figure}

Figure~\ref{fig:uncertainties} summarizes the separately refitted
experimental, PDF, scale, binding-scale, and source-regulator variations.  The
narrow minima in the lower panel are pivot points where a refitted curve crosses
the central result, not physical zeros of an uncertainty.

The principal PDF-scale envelope uses $Q=5$, 10, and 100~GeV.  Every scale
choice is refitted to the same threshold data.  This envelope probes the PDF-evolution-scale dependence at fixed
$\alpha_s=0.3$ in the Coulombic coefficient.  CT18 values at $Q=3.1$ and 2~GeV
are retained as low-scale diagnostics but are not included in the principal
envelope.

The Coulombic binding scale is varied over
$\epsilon_0=0.16$, 0.25, 0.40, and 0.60~GeV.  Its refitted envelope reaches up
to about 28\% at the two lowest measured energies and falls to about 2\% or
less near the upper threshold points.  The resulting spread is treated as a discrete binding-scale model
envelope.  The OPE source
regulator is varied through $W_{\rm src,off}=6$, 8, 10, and 12~GeV with
$W_{\rm src,on}$ fixed.  It changes the six fitted predictions by at most
4.0\%, and by at most 13.2\% over the full $4.04$--$4.80$~GeV grid, including
the unmeasured region below the first point.

When $\epsilon_0$ is held fixed, $m_Q$ and $\alpha_s$ enter the implemented
kernel only through $1/(\alpha_s m_Q^2)$.  Since $c_{\rm OPE}$ is fitted,
independent variations are exact reparameterizations,
\begin{equation}
 c_{\rm OPE}\propto\alpha_s m_Q^2,\qquad
 C_{\rm sub}=\frac{a_{\rm sub}}{c_{\rm OPE}},\qquad
 N_{\rm OPE}=c_{\rm OPE}^2,
\end{equation}
and leave the extrapolated-intercept prediction invariant at machine precision.  This statement
does not apply to a constrained Coulombic model that ties
$\epsilon_0$, $m_Q$, and $\alpha_s$ together.

A conditional Asimov study quantifies the precision required to identify the
TMC shape after refitting both amplitude parameters.  Hypothetical exact-TMC
forward-intercept constraints were added at $W=4.045$, 4.08, 4.12, 4.20,
4.30, and 4.50~GeV, and both hypotheses were refitted with the present sample.
Effective point-to-point uncertainties of 1\%, 0.5\%, and 0.25\% give
$\Delta\chi^2\simeq1.84$, 7.41, and 29.7, respectively.  A common 0--2\%
normalization uncertainty has negligible impact because the discrimination is
driven by curvature.  Within the Coulombic leading-twist model, sub-percent
precision across a broad near-threshold interval is therefore required; the
forecast does not include higher-order or higher-twist ambiguity.

The numerical implementation was cross-checked by comparing nuisance and
covariance formulations, verifying covariance positivity, and testing the
direct convolution, target-mass identities, principal-value integration, and
minimizations independently.  The Peskin-consistent convolution in the
transformed integration variable agrees with an independent logarithmic-$x$
quadrature to $7.4\times10^{-8}$ relative.  The compact-source dispersion grid
agrees with an independent interval-by-interval principal-value evaluation to
$3.8\times10^{-7}$, and the central member of the full uncertainty calculation
reproduces the dedicated threshold fit to $1.9\times10^{-7}$.  All 255
PDF-member optimizations converge.

The uncertainty hierarchy in the measured threshold domain is therefore:
experimental profile uncertainty first; Coulombic binding-scale dependence
next; OPE source-regulator and PDF-scale dependence smaller; and central-set
spread and internal PDF-member uncertainty smallest.  No single total theory
error is formed by adding discrete model envelopes in quadrature.  Genuine
higher-twist operators, higher chromoelectric multipoles, non-Coulombic
finite-size effects, and a fixed-order matched NLO coefficient remain outside
the present calculation.

Existing nonforward differential measurements are important for controlling the
forward-intercept extrapolations and for future amplitude studies in the gap, but they do not by
themselves amplify the target-mass signal within a factorized $t$-slope model.
If the finite-$t$ cross section is written phenomenologically as a common
$t$-dependent form factor multiplying the energy-dependent forward amplitude,
the exact-TMC/no-TMC ratio is unchanged at fixed $t$; additional $t$ bins then
mostly improve statistics and normalization control rather than creating a new
TMC discriminator. Existing nonforward differential measurements are important for controlling
the forward-intercept extrapolations and for future amplitude studies in the
gap, but they do not by themselves amplify the target-mass signal within a
factorized $t$-slope model.  If the finite-$t$ cross section is represented by
a common $t$-dependent form factor multiplying the energy-dependent forward
amplitude, the exact-TMC/no-TMC ratio is unchanged at fixed $t$; additional
$t$ bins then primarily improve statistical and normalization control.  A
finite-$t$ target-mass prediction requires a nonforward OPE or GPD
formulation in which generalized form factors, skewness, and
momentum-transfer corrections are treated consistently.

\section{Conclusions}

Using the Peskin-consistent gluon measure and the exact physical-domain
trace-resummed mapping of Ref.~\cite{Syamtomov:2026tmc}, we evaluate the
leading-twist Coulombic OPE baseline for the forward $J/\psi N$ amplitude
$\mathcal{M}_{\psi N}(\lambda)$.  The published $t$-extrapolated forward
intercepts are related to this amplitude through the VMD normalization of
Eq.~\eqref{eq:vmd}.  The raw target-mass suppression is approximately 5.0--5.5\% near 5~GeV and
1.2--1.3\% at 6~GeV, and falls below one percent for all four central PDF sets
above about 6.4~GeV.  The absorptive amplitude has a finite incoming-endpoint
limit, while the corresponding dispersive real part has a logarithmic cusp;
the numerical comparison is therefore made in the resolved region above the
exact endpoint.

The six GlueX and CLAS12 extrapolated forward intercepts are compatible with
the OPE-generated amplitude dependence after an overall normalization and subtraction
constant are fitted.  They do not, however, identify the exact target-mass
correction: the independently refitted exact-TMC and no-TMC descriptions differ
by only $\Delta\chi^2=0.021$.  A two-parameter empirical exponential gives an
equally good fit, with information-criterion differences below 0.3.  The data
therefore establish compatibility, not preference for a unique threshold
shape.

The subtraction constant is also non-identifiable separately from the overall
amplitude scale.  The point $C_{\rm sub}=0$ lies close to the profile minimum,
with $\Delta\chi^2=0.033$.  The robust threshold result is the model-dependent amplitude modulus implied
by the fitted intercept curve and its profile uncertainty, not a precise
numerical subtraction constant or a directly measured phase.

The HERA measurements independently determine the empirical high-energy trend,
$N_{90}=76.6\pm2.3$~nb and $\delta=0.695\pm0.023$, within the nominal
covariance construction.  ALICE and LHCb proton-target extractions are broadly
consistent with its continuation but do not select among modern small-$x$
mechanisms.  The selected direct-proton data leave a gap from
about 4.73 to 20~GeV.  Since the threshold and HERA likelihoods factorize, no
transition location, width, or phase evolution can be inferred from them.  The
article therefore constrains two separated regimes and identifies an
unresolved crossover rather than measuring an OPE--Regge transition scale.

After refitting, internal PDF-member uncertainties remain below 0.9\% in the
measured threshold range, while the dominant tested model dependence is the
Coulombic binding scale.  A conditional sensitivity study shows that even a
set of six new points over $W=4.045$--$4.50$~GeV would require sub-percent
point-to-point precision to distinguish exact TMC from no TMC once the two
amplitude parameters are refitted.  Measurements throughout the $4.7$--$20$~GeV interval, together with
controlled finite-$t$ dependence and correlated normalizations, are required
before the forward-amplitude crossover itself can be studied quantitatively.
A physical near-threshold analysis at $t=t_{\min}$ would additionally require
a genuinely nonforward OPE/GPD treatment.

\paragraph{Acknowledgments} I thank Christian Otto for inspiring discussions.

\end{document}